\documentclass[aps,pre,twocolumn,showpacs,superscriptaddress,amsmath,amssymb]{revtex4}

\usepackage{graphicx}
\usepackage{dcolumn}
\usepackage{bm}
\usepackage{color}
\usepackage{multirow}
\usepackage{textcomp}
\usepackage{umoline}

\begin{document}

\title{Finite-size scaling in random $K$-satisfiability problems}
\author{Sang Hoon Lee}
\altaffiliation{Present address: IceLab, Department of Physics,
Ume{\aa} University, 901 87 Ume{\aa}, Sweden.}
\affiliation{Department of Physics, Korea Advanced Institute
of Science and Technology, Daejeon 305-701, Korea}

\author{Meesoon Ha}
\email[Corresponding author; ]{msha@kaist.ac.kr}
\affiliation{Department of Physics, Korea Advanced Institute
of Science and Technology, Daejeon 305-701, Korea}

\author{Chanil Jeon}
\affiliation{Department of Physics, Korea Advanced Institute
of Science and Technology, Daejeon 305-701, Korea}

\author{Hawoong Jeong}
\affiliation{Department of Physics, Korea Advanced Institute
of Science and Technology, Daejeon 305-701, Korea}
\affiliation{Institute for the BioCentury, Korea Advanced
Institute of Science and Technology, Daejeon 305-701, Korea}

\date{\today}

\begin{abstract}
We provide a comprehensive view of various phase transitions
in random $K$-satisfiability problems solved
by stochastic-local-search algorithms.
In particular, we focus on the finite-size scaling (FSS) exponent,
which is mathematically important and practically useful in analyzing
finite systems. Using the FSS theory of nonequilibrium
absorbing phase transitions, we show that
the density of unsatisfied clauses clearly indicates
the transition from the solvable (absorbing) phase
to the unsolvable (active) phase
as varying the noise parameter and the density of constraints.
Based on the solution clustering (percolation-type) argument,
we conjecture two possible values of the FSS exponent,
which are confirmed reasonably well
in numerical simulations for $2\le K \le 3$.
\end{abstract}



\maketitle

\section{introduction}
The $K$-satisfiability problem ($K$-SAT) is well known
as nondeterministic polynomial-time (NP) complete when $K\ge 3$.
It is the decision problem of whether an instance of Boolean variables
can be satisfied by variable assignments. The instance is the conjunction ($\wedge$) of clauses
and each clause is the disjunction ($\vee$) of $K$ numbers of variables (or negations).
Determining the $K$-SAT solvability within reasonable computational time
is one of principal unsolved problems
in computer science~\cite{Books}. Moreover, it is fundamentally important,
connected to many applications. Substantial progress of such constraint satisfaction problems
(CSPs) has been achieved~\cite{Books,other-3SAT,Kirkpatrick1994,MWDD-FSS,
Mezard2002,PNAS2007} by either numerical or analytical techniques.

Since the pioneering work for critical behaviors in the random $K$-SAT
by Kirkpatrick and Selman~\cite{Kirkpatrick1994},
mathematics and physics communities have paid attention to structural phase transitions
in the solution space, their scaling behaviors,
and the exact locations of transition points in the thermodynamic limit.
An instance of the $K$-SAT can be interpreted as a $K$-spin interacting system
in statistical physics and its solution as the ground state of the Hamiltonian
for the corresponding spin system. Based on the interpretation,
there are many suggestions for deeper connection
between the criticality in the spin-glass theory
and the intractability of the NP complete problem as well as
many conjectures from both fields by trial and error
regarding computational hardness.

However, few systematic tests of critical behaviors were presented
in the context of finite-size scaling (FSS)~\cite{Kirkpatrick1994,MWDD-FSS}.
In particular, discussions about the FSS exponent
and the transition nature are rare due to the difficulty in finding exact locations
of various transitions in the thermodynamic limit
using finite systems, except for $K=2$ where all the transitions occur
at the same location
as a continuous percolation transition of the solution space.

There are various solving techniques of CSPs available.
For large unstructured CSPs, one can solve
by either general-purpose deterministic algorithms,
e.g., Davis-Putnam-Logemann-Loveland (DPLL)~\cite{DPLL}
(or more tailored message passing algorithms
such belief and survey propagation~\cite{Mezard2002}),
or stochastic-local-search (SLS) algorithms that are generally competitive
for large and least-structured CSPs, e.g., the random $K$-SAT.
In SLS algorithms, assigned values to variables are successively flipped,
based on the local information of algorithmic details.
Starting with the celebrated simulated annealing algorithm
by Kirkpatrick {\it et al.}~\cite{Kirkpatrick1983},
several focused SLS algorithms have been developed:
RandomWalkSAT~\cite{Papadimitriou1991}, WalkSAT~\cite{Selman1996},
focused Metropolis search (FMS)~\cite{Seitz2005},
and average SAT (ASAT)~\cite{Ardelius2006}.

All the solving techniques, however,
have difficulties in approaching
a sharp change of the ensemble for random CSPs,
namely a ``phase transition". Deterministic one,
in spite of its exactness, suffers from severely limited system sizes,
while stochastic one is able to deal with much larger system sizes
but their results are less accurate than the former due to fluctuations and
some ambiguity caused by the limited simulation time~\cite{Barthel2002+}.

In this paper, we propose a systematic method to analyze data
obtained from finite systems, which can resolve
numerical accuracy issues from the limited system size,
the method of sampling, and the computational time.
It is based on the FSS analysis of
nonequilibrium absorbing phase transitions (APTs)~\cite{MarroBook}.
We employ it to characterize critical behaviors
of the transition from the solvable [(SOL) absorbing] phase to the unsolvable
[(UNSOL) active] phase in the random $K$-SAT,
in terms of the density of unsatisfied (UNSAT) clauses
as an indicator and a solution as an absorbing state.

This paper is organized as follows. In Sec.~\ref{K-SAT+ASAT},
we describe the random $K$-SAT and explain how to explore it
by ASAT heuristic. In Sec.~\ref{FSS}, we suggest relevant physical quantities,
and discuss the main idea of FSS ansatz in perspective of nonequilibrium APTs.
We also argue scaling properties
near and at dynamic SOL/UNSOL phase transitions,
which are numerically confirmed well in Sec.~\ref{numerics}.
Finally, we conclude this paper in Sec.~\ref{conclusion}
with the summary of the main results and some remarks.

\section{Random $K$-SAT and ASAT Heuristic}
\label{K-SAT+ASAT}

The Boolean expression of an {\em instance} $F$ in the random $K$-SAT is written as
$F = [ C_1~\wedge~C_2~\wedge~\cdots~\wedge~C_M],$
where each {\em clause} $C_i$ is given by
$C_i = (y_{i1}~\vee~y_{i2}~\vee~\cdots ~ \vee ~ y_{iK} ),$
and each value of $y_{ij}$ is randomly assigned from the set
$\{ x_1,\ \neg{x}_1,...,\ x_N,\ \neg{x}_N\}$
of $2N$ Boolean variables (themselves and their negation).
The above conjunctive norm form of $F$ can be also expressed as
a bipartite network (factor graph) form, too.
The density of constraints ($\alpha \equiv M / N$)
plays the role of a control parameter in the random $K$-SAT
since it can determine the satisfiability~\cite{Kirkpatrick1994}
and the average solving time of algorithms~\cite{Mitchell1992}.
As $\alpha$ increases, it gets harder to find the SAT configuration of variables,
and eventually the solution does not exist for too large $\alpha$ values.
At least one threshold, therefore, must exist between the SAT and UNSAT phases.

Using the most recently developed SLS algorithm, ASAT~\cite{Ardelius2006},
we systematically show how to find such a threshold value
from numerical data of finite systems, denoted as $\alpha_c$
in the thermodynamic limit, as well as critical exponents. They correspond to
the solvability transition point and its critical behaviors,
very similar to those of nonequilibrium APTs. Among lots of algorithms,
ASAT deserves to be considered the representative case of our new FSS analysis
because it is not only the most efficient focused SLS heuristic
but also the simplest variant of well-known algorithms with the specific value
of the noise parameter of ASAT, e.g., RandomWalkSAT.

Since our main interest is the minimal model study of the random $K$-SAT,
we here present only the results of ASAT and its limiting case, RandomWalkSAT,
but our analysis techniques can be easily applied to any other algorithms
(partially tested in~\cite{OurFullPaper}).
For those who are interested in the graph coloring problem
($Q$-COL), WalkCOL in~\cite{Zdeborova2007}
would be the best to be tested by the same FSS analysis as what we do.
It is because WalkCOL is the exact adaptation of ASAT in the random $Q$-COL.

We explore the random $K$-SAT by ASAT as follows:
choose a clause at random among the set of UNSAT clauses
and then randomly try flipping one assigned value
out of $K$ variables in the chosen clause. The trial flip is accepted
with certainty unless the total number of UNSAT clauses,
$M_u$
increases, or with a probability $p$ (noise parameter)
if $M_u$ increases.
Whether each trial flip is accepted or not, time is incremented
by $\Delta t$. In general, one takes $\Delta t=1/N$ (or $1/M$)
where $N$ is the total number of variables and $M=\alpha N$,
so that a unit time interval (Monte Carlo step) corresponds
to one trial flip per variable on average.
However, our choice is restricted to UNSAT clauses
only in order to improve the simulation efficiency of ASAT
(by definition, it is a focused heuristic),
such that $\Delta t=1/M_u(t)$. Here $M_u(t)$ is
the total number of UNSAT clauses at time $t$.
The simulation is terminated either if a solution is
found or if the given instance is not solved yet
until the maximal time, $T_\textrm{max}$.

\section{Finite-Size Scaling Ansatz}
\label{FSS}

So far, critical behaviors near the SOL-UNSOL transition have been discussed
in terms of the fraction of solved/successful samples,
$P_s(\alpha, N)$, and the FSS exponent $\bar{\nu}$
that determines the FSS width of a continuous phase transition
as $|\epsilon|N^{1/{\bar{\nu}}}$, where $\epsilon=(\alpha-\alpha_c)/\alpha_c$.
This is based on the fact that there is some diverging correlation volume,
$\xi_\textrm{v}\sim |\epsilon|^{-\bar{\nu}}$
($\xi_\textrm{v}=N$ in finite systems at $\epsilon=0$),
like the diverging correlation length as $\xi\sim |\epsilon|^{-\nu}$
($\xi=L$ with $L=N^{1/d}$ in $d$-dimensional finite lattices at $\epsilon=0$).
One can find the detailed discussion of $\bar{\nu}$
for non-regular lattice types, complex networks,
in~\cite{HHP2007}.

Our FSS analysis in the random $K$-SAT follows the postulate of
a diverging dynamical correlation volume, $\xi_{\textrm v}$,
at the solvability transition whose physical manifestation is
the presence of dynamical heterogeneities with infinitely
many solution states. Using the analogy of the FSS concept in the static simulations
of nonequilibrium APTs, we measure two more physical quantities,
(besides the solved-sample fraction, $P_s$), playing roles
as good and independent indicators in SLS algorithmic phase transitions:
the solving time and the density of UNSAT clauses.

The solving time, $\tau$, can be determined in two ways from $P_s(\alpha,N,t)$
for $t\le T_\textrm{max}$: (1) $\tau_{_\textrm{H}}(\alpha,N)=t^*$
when $P_s(\alpha,N,t^*)=1/2$, corresponding to the median value of
the solution time set. (2) $[\tau(\alpha,N)]$, where [$\cdot$] denotes
an average restricted to SOL trials before $T_\textrm{max}$
out of all trial samples. Since both
are well defined in the SOL phase ($\epsilon < 0$),
they indicate the transition into the UNSOL phase ($\epsilon > 0$)
for $N\gg \xi_\textrm{v}$ as $\tau \sim |\epsilon|^{-\nu_{\parallel}}$,
like the relaxation time in APTs.
Incorporating the size dependence generally yields
$\tau(\alpha,N)=N^{\bar{z}}h(\epsilon N^{1/\bar{\nu}})$,
where $h(x)\sim x^{-\nu_{\parallel}}$ for large $x$ and
$\tau\sim N^{\bar{z}}$ at $\epsilon=0$
with $\bar{z}=\nu_{\parallel}/\bar{\nu}$.
In the SOL phase, $\tau$ approaches a constant as $N\to\infty$,
while in the UNSOL phase it grows exponentially with $N$.
It is noted that we present $\tau_{_\textrm{H}}$ only.

The density of UNSAT clauses, $\rho_u(\equiv M_u/N)$,
plays a role of another good indicator in the solvability transition of
the random $K$-SAT, namely, ``active clause''
density as if the order parameter of APTs.
In applying FSS to its critical behaviors,
one should notice that the true stationary state of
a finite system is only the SOL state.
To learn about the UNSOL state from algorithm tests,
one should investigate the quasistationary state
describing the statistical properties of
UNSOL trials with some initial transient,
and determine such quasistationary properties
from averages over UNSOL representatives
out of a large independent trial set with random initial conditions.
After the initial transient (depending on both $\alpha$ and $N$),
$\langle\rho_u(\alpha,N,t)\rangle=\langle \rho_u(\alpha,N,t)\rangle_\textrm{all} /
\{1-P_s(\alpha,N,t)\},$ which gets saturated to $\langle\tilde{\rho_u}(\alpha,N)\rangle$
for $t\gg N^{\bar{z}}$. Here $\langle\cdot\rangle$ corresponds to an average restricted
to UNSOL trials and $\langle\tilde{\cdot}\rangle$
to an average of saturated steady values.

Near the transition for small $|\epsilon|$ and large $N$,
the survival UNSAT density is written in the FSS form,
$\langle\tilde{\rho_u}(\alpha,N)\rangle
=N^{-\theta}g(\epsilon N^{1/\bar{\nu}})$,
where $g(x)\sim x^{\theta\bar{\nu}}$ for $x\gg 1$ and $N\gg\xi_\textrm{v}$.
In the SOL phase, it trivially scales as
 $\langle\tilde{\rho_u}(\alpha,N)\rangle\sim N^{-1}$,
so $g(x)\sim |x|^{\bar{\nu}(1+\theta)}$ for negatively large $x$.
The $\alpha_c$ value may also be found by examining its $N$-dependence
as $\langle\tilde{\rho_u}(\alpha_c,N)\rangle\sim N^{-\theta}$
since in the SOL phase, it falls off as $N^{-1}$,
while in the UNSOL phase, it approaches an $\alpha$-dependent value.

Finally, we explain the dynamic scaling of $\langle\rho_u(N,t)\rangle$,
averaged over survival trials at $\alpha_c$ with random initial configurations,
where the time dependence only involves the ratio $t/N^{\bar{z}}$,
so that $\langle\rho_u(N,t)\rangle\sim N^{-\theta}f(t/N^{\bar{z}})$.
It is, however, hard to observe the saturated regime for $t\gg N^{\bar{z}}
(\xi_\textrm{v}\sim t^{1/\bar{z}})$ as $N$ increases,
so it is better to focus on the temporal decay regime
for $t\ll N^{\bar{z}},\ \textrm{i.e.,}\ \xi_\textrm{v}\ll N$ with the largest $N$ value
one can test, which enables to determine both $\alpha_c$ and $\delta(=\theta/\bar{z})$
at the same time with the pretty good accuracy.
This is why $\langle\rho_u(N,t)\rangle\sim t^{-\delta}$ should be first investigated
without any assumption of the $\theta$ value:
$\langle\rho_u(N,t)\rangle=t^{-\delta}F(t/N^{\bar{z}})$,
where $F(x)=\textrm{constant}$ for $x\ll 1$ and $F(x)\sim x^{\delta}$ for $x\gg 1$.
\begin{figure}[t]
\begin{tabular}{c}
\includegraphics[width=0.75\columnwidth]{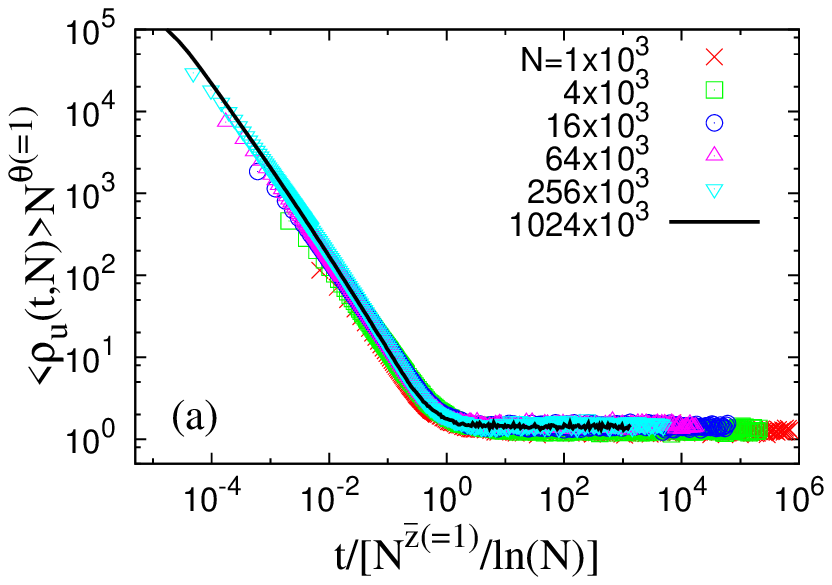}\\
\includegraphics[width=0.75\columnwidth]{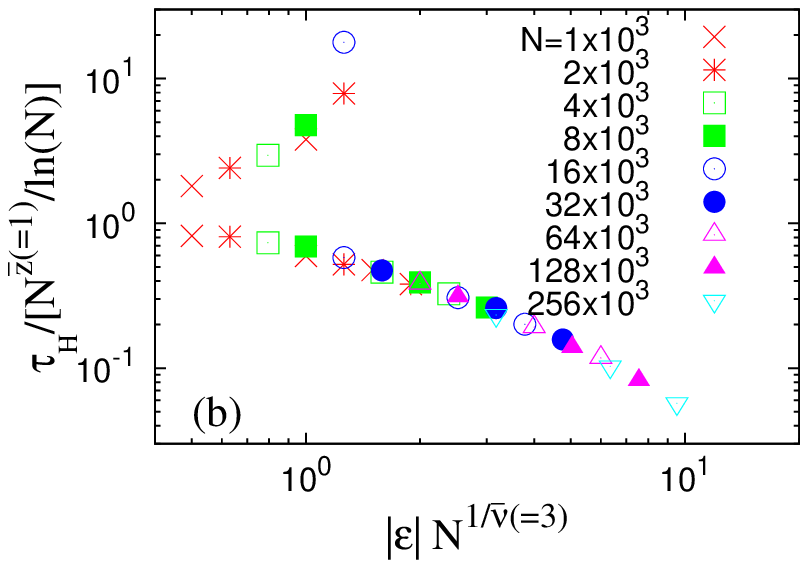}
\end{tabular}
\caption{(Color online) FSS for 2-SAT by ASAT with $p=1/2$,
where logarithmic corrections to scalings are found as
(a) $\langle\rho_u(N,t)\rangle N^{\theta}=f(t/[N^{\bar{z}}/\ln(N)])$ and
(b) $\tau_{_\textrm{H}}(\alpha,N)/[N^{\bar{z}}/\ln(N)]=h(|\epsilon| N^{1/\bar{\nu}})$
with $\epsilon=(\alpha-\alpha_c)/\alpha_c$.
 For the convenience,
the same symbols and lines are taken to the same system sizes
in Figs.~\ref{fig2-RandomWalk} and~\ref{fig3-ASAT_optimal}.}
\label{fig1_2-SAT}
\end{figure}
\begin{figure}[t]
\begin{tabular}{c}
\includegraphics[width=0.75\columnwidth]{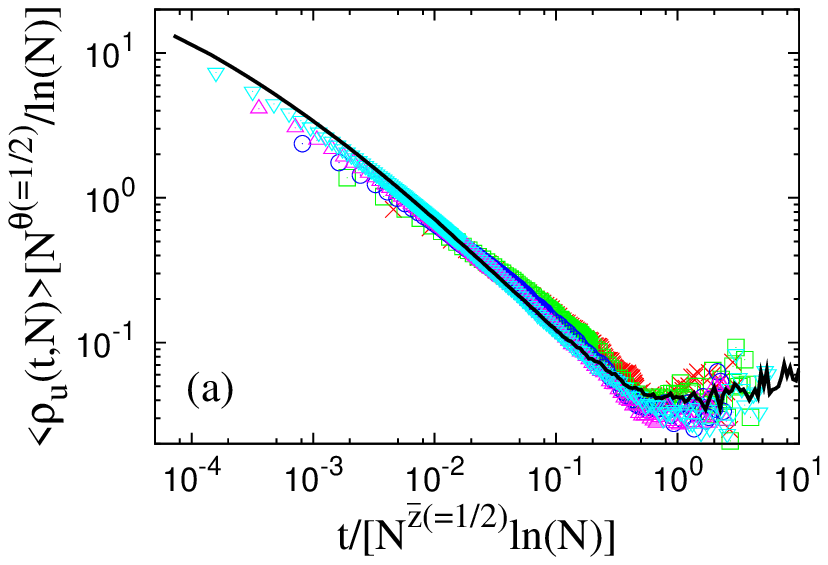}\\
\includegraphics[width=0.75\columnwidth]{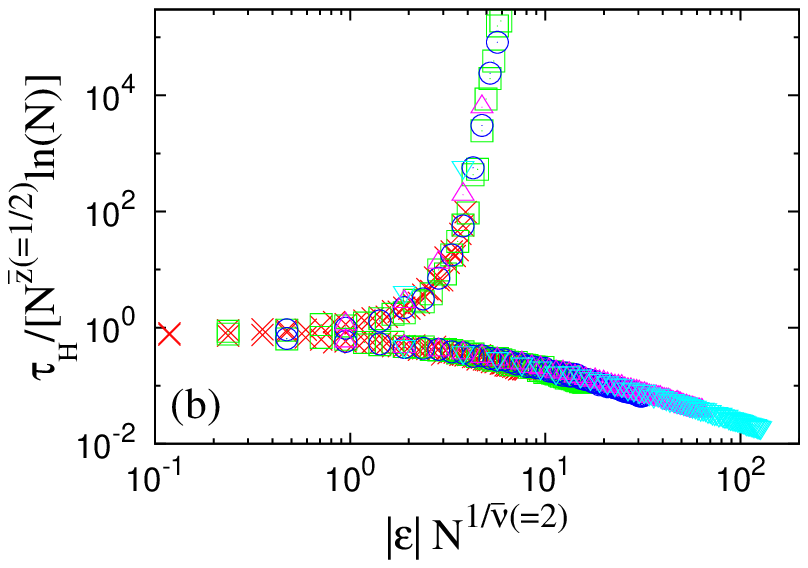}
\end{tabular}
\caption{(Color online) FSS for 3-SAT by the limiting case of ASAT ($p=1$), RandomWalkSAT,
with logarithmic corrections to scalings as
(a) $\langle\rho_u(N,t)\rangle[N^{\theta}/\ln(N)]=f(t/[N^{\bar{z}}\ln(N)])$ and
(b) $\tau_{_\textrm{H}}(\alpha,N)/[N^{\bar{z}}\ln(N)]=h(|\epsilon| N^{1/\bar{\nu}})$.
The same symbols and lines are taken to the same system sizes
as in Fig.~\ref{fig1_2-SAT}.}
\label{fig2-RandomWalk}
\end{figure}

\section{Numerical Results}
\label{numerics}
\begin{figure}[t]
\begin{tabular}{c}
\includegraphics[width=0.75\columnwidth]{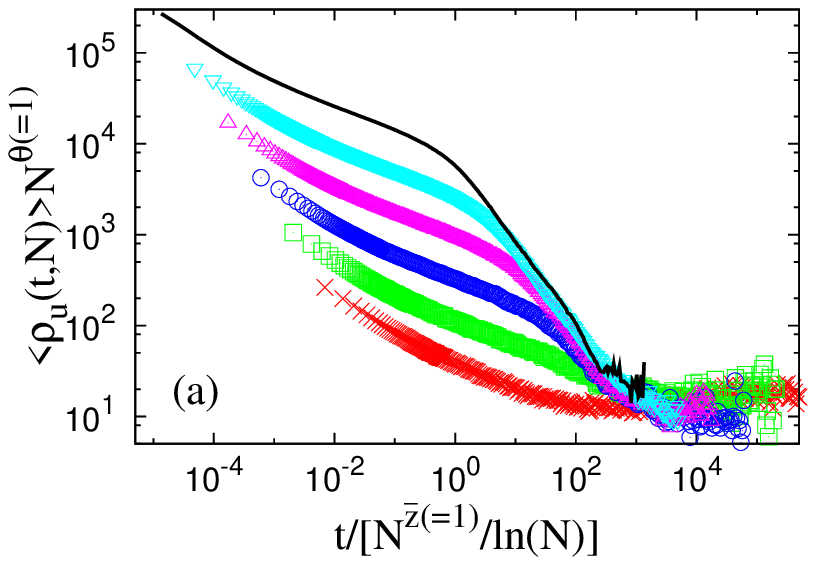}\\
\includegraphics[width=0.75\columnwidth]{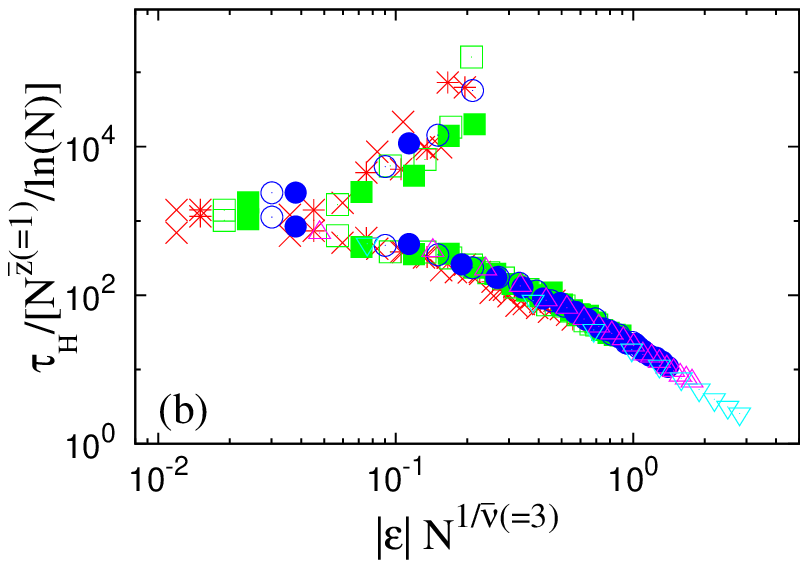}
\end{tabular}
\caption{(Color online) FSS for 3-SAT by the optimized ASAT ($p=p_\textrm{opt}\simeq0.21$)
with logarithmic corrections to scalings as
(a) $\langle\rho_u(N,t)\rangle N^{\theta}=f(t/[N^{\bar{z}}/\ln(N)])$
and (b) $\tau_{_\textrm{H}}(\alpha,N)/[N^{\bar{z}}/\ln(N)]=h(|\epsilon| N^{1/\bar{\nu}})$.
The same symbols and lines are taken to the same system sizes
as in Fig.~\ref{fig1_2-SAT}.}
\label{fig3-ASAT_optimal}
\end{figure}
We now present scaling properties tested
for $2\le K\le 3$ by ASAT with the noise parameter $p$,
where we set $T_\textrm{max} = 10^8$ and test at most $10^3$
($5\times 10^2$) samples for 2-SAT (3-SAT).
The values of $p$ are chosen as follows:
SLS algorithms may have the {\em optimal $p$ value}
that exists between too less noise to prevent the escape of the system
from local energy minima and too much fluctuations. By definition,
an optimized algorithm ($p=p_\textrm{opt}$) finds solutions
with the fastest solving time up to the largest $\alpha$ value.
It was reported that ASAT for 3-SAT, $p_\textrm{opt}\simeq 0.21$,
allowing to find solutions up to $\alpha_\textrm{lin}\simeq 4.21$
where the number of flipping variables is linearly proportional
to $N$ until a solution is found~\cite{Ardelius2006}.
In contrast, there are no optimal $p$ values for 2-SAT.
It seems to be because all the transitions occur at $\alpha_c=1$
as the mean-field (MF) {\em percolation} transition with $\bar{\nu}=3$.
Such a conjecture has been confirmed by the same FSS test~\cite{OurFullPaper}
in $(2+X)$-SAT with $X\in[0,1]$ (well-discussed in~\cite{MWDD-FSS}).
Up to a specific $X^*$ value ($X^*=2/5$),
it behaves as if 2-SAT without the complexity issue of the solution space.
Here $X$ is the probability for 3-SAT clauses in an instance.

Figure~\ref{fig1_2-SAT} shows FSS tests for 2-SAT by ASAT with $p=1/2$,
where critical exponents are obtained from $\tau$ and $\rho_u$ as varying
$\alpha,\ N,\ \textrm{and}\ t$. In particular, we indicate the precise
$\alpha_c$ location as $\alpha_c=1.00(2)$ with $\delta=1.0$
using the plateau and inflection-point analysis of effective exponent plots
for various system sizes (not shown here). Through the conventional FSS analysis,
we obtain $\theta=\bar{z}=1.0$ and $\bar{\nu}=3.0$,
where $\delta= \theta/\bar{z}$ is also checked within error bars.
Note that logarithmic correction to scalings exist
as $\tau_{_\textrm{H}}\sim N^{\bar{z}}/\ln(N)$, stemming from
the presence of quenched disorder in finite CSPs.
Scaling behaviors of 2-SAT,
including $(2+X)$-SAT for $X\le X^*$,
do not depend on $p$ in ASAT, indeed, and
even in the limiting case of ASAT, RandomWalkSAT ($p=1$), as well.
In spite of the well-known results of 2-SAT,
its detailed scaling properties have rarely been
checked systematically for finite systems. Thus, our FSS analysis in 2-SAT
could be a prototype of further applications, including
our test in 3-SAT where we find an interesting result that critical behaviors
in RandomWalkSAT are quite different from those in the optimized ASAT
with $p=p_\textrm{opt}=0.21$ using the same analysis as Fig.~\ref{fig1_2-SAT}.

Figure~\ref{fig2-RandomWalk} shows FSS tests for 3-SAT by RandomWalkSAT
at $\alpha_c=2.670(5)$ with $\theta=\bar{z}=0.50$ and $\bar{\nu}=2.0$,
where the precise location of $\alpha_c$ is first identified with $\delta=1.0$.
These results are exactly the same as those
in the MF {\em directed percolation} (DP) transition
with infinitely many absorbing states~\cite{MarroBook},
within the SAT phase of 3-SAT, even though there are logarithmic
corrections to scalings again: $\langle\rho_u(t)\rangle\sim [\ln(t)]^{0.25}/t,\
\langle\tilde{\rho}_u(N)\rangle\sim \ln(N)/\sqrt{N},\ \textrm{and}\
\tau_{_\textrm{H}}\sim \sqrt{N}\ln(N)$, respectively.

However, the optimized ASAT for 3-SAT ($p_\textrm{opt} = 0.21$)
exhibits totally different scaling behaviors from those in RandomWalkSAT.
It is because its transition is located well below
the clustering and condensation transition threshold,
$\alpha_c<\alpha_d$, where $\alpha_d\simeq 3.86$
in~\cite{PNAS2007} (related to the MF percolation transition
of solutions for 3-SAT), while that of the optimized ASAT
is much above $\alpha_d$ and rather close to $\alpha_s$ (the SAT-UNSAT threshold).

At the first sight of Fig.~\ref{fig3-ASAT_optimal}(a),
the FSS collapse of $\langle \rho_u (t, N) \rangle$ does not seems
to be good with $\theta=\bar{z}=1.0$ due to $N$-independent initial transient
before the true scaling regime. We find a transition occurring at $\alpha_c=4.185(5)$
and $\delta=1.0$ accompanying with logarithmic corrections to scalings.
After the initial transient, the true scaling regime becomes clearly extended as $N$ increases,
while in small system sizes, $O(10^3)$, such a regime is absent.
Figure~\ref{fig3-ASAT_optimal}(b) shows that $\bar{\nu}=3.0$ and $\bar{z}=1.0$
with logarithmic corrections to scalings (same as Fig.~\ref{fig1_2-SAT}).
We are also aware at $\alpha^*=4.26$ (very near $\alpha_s$)
of a nontrivial power-law decay exponent,
$\delta\simeq 0.20(5)$ (or logarithmic scaling)~\cite{OurFullPaper},
but it is not relevant to our current work, so not shown here.
Relevant numerical results are summarized in Table~\ref{table1}.
\begin{table}[t]
\caption{SOL-UNSOL threshold and critical exponents
for 2-SAT and 3-SAT by ASAT with the noise parameter $p$.}
\label{table1}
\begin{tabular}{ccccccc}
\hline\hline
 &$p$&$\alpha_c$&$\delta$&$\bar{z}$&$\theta$&$\bar{\nu}$\\
\hline
2-SAT&~Any $p(>0)$~&1.00(2)&1.0&1.0&1.0&3.0\\
\hline
3-SAT&$~1.00~$&2.670(5)&1.0&0.5&0.5&2.0\\
     &$~0.21~$&4.185(5)&1.0&1.0&1.0&3.0\\
\hline\hline
\end{tabular}
\end{table}

\section{Summary}
\label{conclusion}

We have analyzed the random $K$-SAT
by the simplest SLS heuristic
in the numerical framework of nonequilibrium APTs.
Two possible values of the FSS exponent ($\bar{\nu}$) in 3-SAT are conjectured:
one is $\bar{\nu}=2$ in the {\it directed percolation} university class
with infinitely many absorbing states~\cite{MarroBook}
if $\alpha_c < \alpha_d$. The other is $\bar{\nu}=3$
in the {\it percolation} university class (same as 2-SAT)
if $\alpha_d<\alpha_c<\alpha_s$,
where $\alpha_d$ is the condensation and clustering threshold
and $\alpha_s$ is the SAT-UNSAT threshold~\cite{Mezard2002,PNAS2007},
which are numerically confirmed with logarithmic corrections to scalings.

In conclusion, we have a few remarks for further studies:
dealing with numerical data in $K$-SAT,
one should know serious finite-size effects
of small systems, $N \lesssim O(10^3)$.
The FSS analysis we tested here would be widely
applicable to test constraint satisfaction problem (CSP) algorithms' performance.
The validity check of our results and methods could be possible in the graph $Q$-coloring problem ($Q$-COL).
Finally, we suggest that it would be interesting to investigate
how the sampling bias of SLS algorithms discussed in~\cite{Mann2010a}
affects our results (already smeared)
in universality perspective of the SOL-UNSOL transition.

\section*{ACKNOWLEDGMENTS}
This work was supported by NAP of Korea Research Council
of Fundamental Science and Technology and by the BK21 project.
S.H.L. and M.H. would acknowledge fruitful discussion
with John Ardelius, Erik Aurell and Mikko Alava,
and the kind hospitality of NORDITA, where the main idea was initiated.
M.H. would appreciate valuable comments by Haijun Zhou, Lenka Zdeborov{\'a},
Alexander K. Hartmann and Stefan Boettcher. H.J. would acknowledge APCTP for generous hospitality. Computation was partially carried out using KIAS supercomputers.


\end{document}